# Adhesive Penetration in Beech Wood

# Part I: Experiments

P. Hass*, F.K. Wittel, M. Mendoza, H.J. Herrmann, and P. Niemz

Institute for Building Materials, ETH Zurich, Schafmattstrasse 6, CH-8093 Zurich.

*phass@ethz.ch

## Abstract

A study with synchrotron radiation X-ray tomographic microscopy (SRXTM) of PUR, PVAC, and UF adhesive bond lines in beech wood, bonded under various growth ring angles is presented. After determining the hardening characteristics of the adhesives, we evaluate the bond line morphologies, and the adhesive penetration into the wood structure. We find distinct bond line imperfections for the different adhesive systems. To describe the adhesive distribution inside the bond line we propose the saturation of the pore space instead of the commonly used maximum penetration depth. The results are the basis for a penetration model of hardening fluids into hardwood (part II).

# Introduction

Modern timber engineering strongly relies on wood adhesive bonding. The majority of highly engineered wood-based products have been developed on the principle of cutting wood into smaller pieces and joining them again by adhesive bonding. In this way, the natural anisotropy of wood can be reduced, bigger dimensions of structural elements become possible and the characteristics of the components can be improved. Even though the bonding of wood elements is a rather simple processing step, the details of the penetration of the hardening adhesives into the porous wood skeleton on several scales are rather complicated. It is strongly influenced by wood factors like wood species, anatomical orientation or surface roughness, adhesive factors such as adhesive type or viscosity and process factors as applied pressure or temperature with a big influence on the bonding performance (Kamke and Lee 2007). Former studies on adhesive penetration are mainly performed using microscopy of cross sections and micro-slides (Kamke and Lee 2007; Sernek et al. 1999; Suchsland 1958). More integral studies were performed using porosimetry (Wang and Yan 2005) or neutron radiography (Niemz et al. 2004). To measure adhesive penetration of cell walls, scanning-thermal microscopy (SThM) was used (Konnerth et al. 2008). In most cases, the penetration behavior of an adhesive is described using the maximum penetration depth, without differentiating between wood species. The differences of the penetration behavior between different wood species is exemplarily demonstrated in Figure 1, where bond lines of the same adhesive are shown in a softwood (Figure 1(a)) and a hardwood (Figure 1(b)).

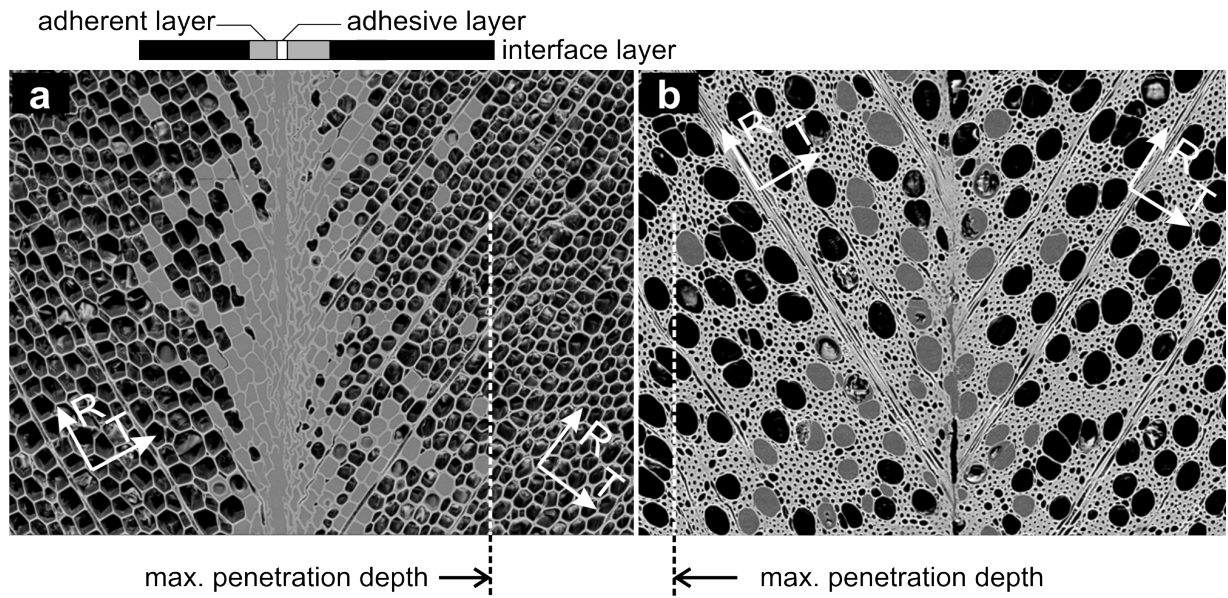

**Figure 1:** PUR Bond lines in spruce (a) and beech wood (b).

In softwood, adhesive filled tracheids are visible as a more or less interconnected zone. Hence, for softwood it seems probable to express the penetration behavior by the maximum penetration depth via a simple trigonometric relation describing the filling of uni-directional tracheids that are cut at the bond line (Suchsland 1958). For hardwood the picture is far more complicated, as many adhesive filled vessels appear to be isolated from the bond line. Here the maximum penetration depth seems to be an arbitrary measure and insufficient to describe the penetration behavior. Hence a different approach has to be found which gives a better characterization of the adhesive penetration, by considering the anatomical structure of the specimen.

In part I of this work a systematic study of adhesive penetration into beech wood (*Fagus sylvatica* L.) samples is described. Specimens are bonded with parallel longitudinal axes under varying growth ring angles using three different adhesive systems with different initial viscosities. To measure the internal distribution of the adhesives, tomographic imaging via synchrotron radiation is employed. From the high resolution 3D-data, qualitative investigations of the bond line are feasible, which allow for example the detection of flaws resulting from the curing process. By image processing, namely image transformation, spatial convolution on 3D data, segmentation and morphologic operations, the adhesive and the pore space are segmented. Thereby it is possible to characterize the adhesive penetration as the saturation of the accessible pore space.

In part II, we propose a hierarchical analytical model for adhesive penetration in hardwood. The first scale describes the capillary transport of a hardening adhesive in a single vessel in time with the possibility of constant diffusion of solvent trough the vessel wall. When the viscosity increases by hardening and/or loss of solvent, the adhesive front slows down and finally stops. Since the result depends on vessel diameters, the results are embedded in a three-dimensional fishing net model with identical network properties like pore-size distribution and connectivity of the vessel network.

## Material and Methods

In order to provide a sound basis for a adhesive penetration model, we need to characterize the wood anatomy, including the pore space (see Hass et al. 2010) and the ray distribution, the viscosity evolution of the adhesive and finally the bond line morphology for verification.

### Adhesive Systems

To get a general idea about the penetration process of adhesives into wood, and the differences between various adhesives, three major adhesive types were investigated. An overview on the adhesive characteristics and their chemical composition is given in (Dunky 2002, Habenicht 2006).

- Urea formaldehyde (UF) is considered to be the most important amino plastic bonding agent in the area of manufactured wood products for internal use without high moisture exposition. Hardening of UF consists of the loss of the solvent (water) and the actual hardening reaction, where the urea reacts with formaldehyde in an acid environment in the form of a poly condensation. The hardened adhesive builds insolvable infusible spatial networks leading to brittle, duroplastic material behavior of the bond line).

- Polyvinyl acetate (PVAC) builds the second most important adhesive group for the furniture industry. The hardening of PVAC is characterized by a physical hardening process, which is initiated by the loss of the water contained in the adhesive dispersion. Under the consolidation pressure applied on the adherents, an adhesive film is formed and repulsive forces between the single PVAC molecules are resolved and the adhesive layer becomes solid.

- One component polyurethane (1K-PUR), which gains increasing importance for load-bearing timber structures and solid wood constructions. Those adhesives consist basically of polyols and isocyanates, which react with the OH-groups of the adherent surfaces with disposal of $CO_2$. To adjust the adhesives viscosity, organic solvents can be used or the cross-linking density of the pre-polymer can be adjusted. Polyurethane pre-polymers only contain the basic ingredients and no additives as for example fillers or defoamers, which will be added in a ready for sale adhesive.

**Hardening**

In a further step, we perform measurements to describe the hardening behavior of the pure adhesives. For UF, we prepare mixtures of 100g UF adhesive powder with 80, 70, 60, and 50 ml of water. After correct mixing, the evolution of viscosity and temperature of the different mixtures is tracked via a rotational viscosimeter (Haake Viscotester VT5R) and a thermometer (Figure 2).

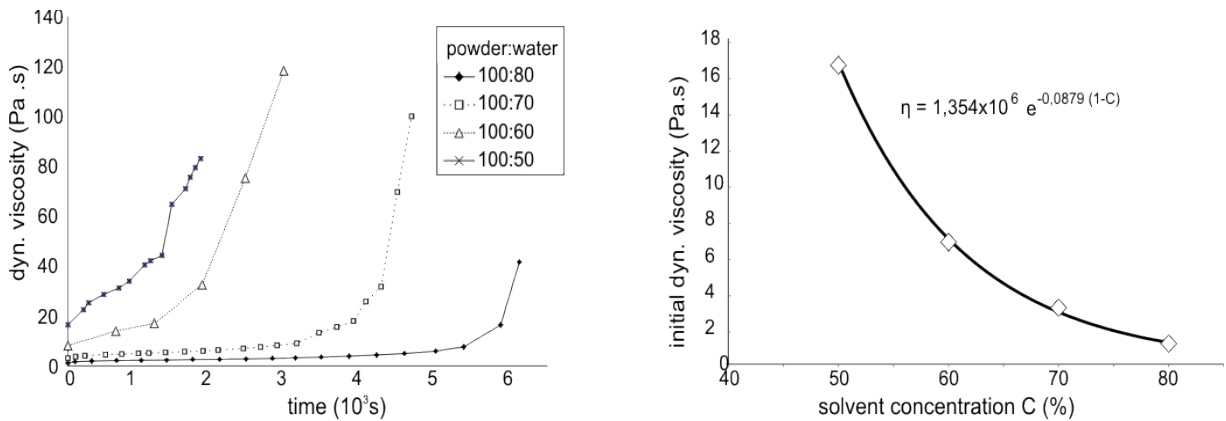

**Figure 2:** (left) Evolution of the dynamic viscosity of hardening UF-adhesives with different solvent content and (right) initial viscosity.

Since the measured temperature changes during hardening stay below 1°C, they are negligible. For PUR and PVAC such measurements are not possible, since PUR becomes foamy during hardening and PVAC builds a skin on the surface of the sample which prevents further loss of water from the mixture. Hence the viscosity stays almost constant under the skin. However the measurements for PVAC can be performed inversely. As the hardening of PVAC depends on the loss of the water the PVAC molecules are dispersed in, the change in solid content and thereby viscosity can be used to describe the hardening process, if the parameters for the water uptake by the wood are known as well. By adding water to the PVAC mixture with the higher initial viscosity, the solid content can be changed until the one of the second PVAC adhesive is reached (Figure 3).

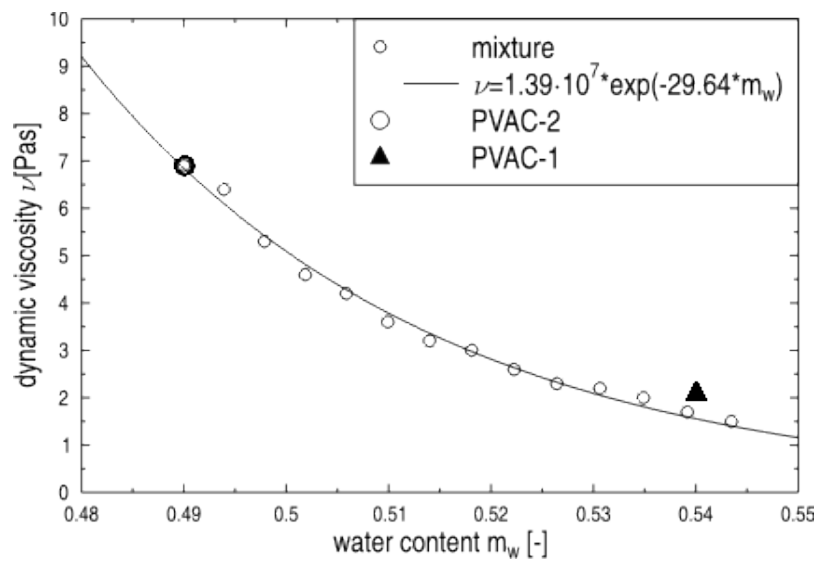

**Figure 3:** Evolution of the dynamic viscosity of PVAC with differing water-solid ratio.

**Bond Line Morphology**

For the investigation of the volumetric distribution of the adhesives inside the wood after hardening, μ-tomographies have been conducted on bonded beech wood samples using synchrotron radiation X-ray tomographic microscopy (SRXTM). The influence of the anatomical direction on the adhesive penetration is covered by bonding the wood under various growth ring angles (GRA) from radial to tangential orientation in 15° steps (Figure 4) keeping the longitudinal axes parallel. To allow a later comparison to the failure behavior and the strength of the bonding, the samples were chosen from lap-shear specimens used in (Hass et al. 2009). From the three PUR-prepolymers used in that investigation only specimens bonded with prepolymer

1 (initial viscosity 1310 mPas) and prepolymer 2 (initial viscosity 5460 mPas) were used. For PVAC and UF, both adhesive variants per system were investigated. This gave a total number of 42 specimens, which were surveyed using synchrotron radiation X-ray tomographic microscopy (SRXTM). The details of the investigation method are well described in (Hass et al. 2010), which is based on the same set of samples as this study. Today SRXTM is a relatively novel technology that does not allow huge sample size studies. Facilities to perform such measurements are scarce and enormous amounts of data have to be processed for measuring, reconstruction and evaluation. By choosing only one sample per adhesive and growth ring angle step, a well-founded statistical investigation on the influence of the GRA is not possible, but as the aim on this survey laid on the major principles of the penetration behavior, we preferred to increase the number of influencing factors than the number of repetitions.

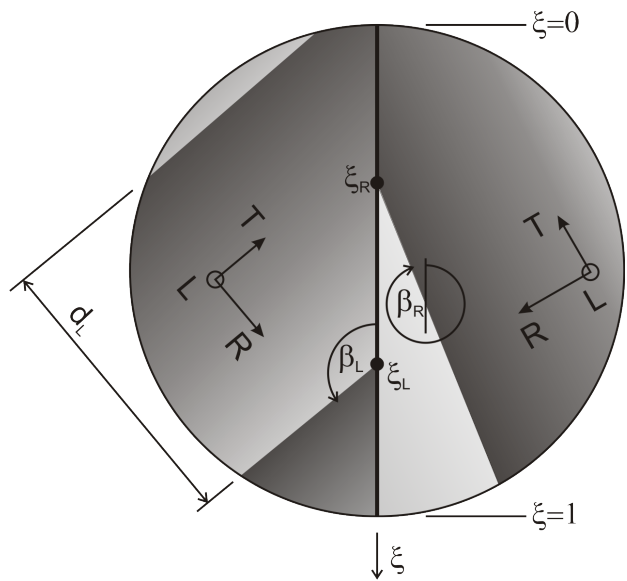

**Figure 4:** Definition of growth ring orientations and positions in local coordinates of a cylindrical specimen. Signs and symbols: R / T / L→ radial / tangential / longitudinal direction; β → growth ring angle; d → growth ring width; ξ → position in bond line; subscriptions: L → left side; R → right side

The μ–CT data consists of a stack of images of cross-sections in the RT-plane averaging a volume of 3.7μm$^3$ by one scalar gray-value. First all images are rotated in the uniform coordinate system (Figure 4) using a bicubic interpolation that leads to a weighted average of pixels in the nearest neighborhood (MatLab 2009). Then the resolution is doubled and gray values are adjusted to match the full intensity range. Due to contrast shifts along the longitudinal axis, the gray value distribution is shifted. At periodic height levels (every 100$^{th}$ slice) gray values of pure adhesive zones are picked to obtain sub-block wise the gray value distribution of the adhesive. These slices are used for the investigation of the adhesive penetration, by analyzing the maximum penetration depth and the adhesive saturation of the available pore space. While the distribution for pores is isolated (see Hass et al. 2010), the ones for adhesives and wood strongly overlap, complicating the segmentation. Note that the addition of a contrast enhancing substance to the adhesive was dismissed for the uncertainties this procedure inherits. For example, (Modzel et al. 2010) added rubidium to increase the X-ray attenuation of their adhesive. Unfortunately the rubidium traveled deeper into the wood than the adhesive, questioning the suitability of this method. Apart from the possible negative effects for the identification of the adhesive, changes of the adhesive properties might occur.

To segment the adhesive in our data, we apply a two step procedure: First an intensity based segmentation using the fitted normal distribution for local thresholding is performed. Since adhesive regions are rather compact, morphologic image processing on binarized images can be applied to filter small segmented zones away from the bond line inside the sub-block. Contour smoothening can be obtained by morphologic opening of blops and finally small holes inside the blobs can be filled, if desired. Additionally small blops without depth can be deleted by a blob analysis on the segmented tomographic data (see Figure 5).

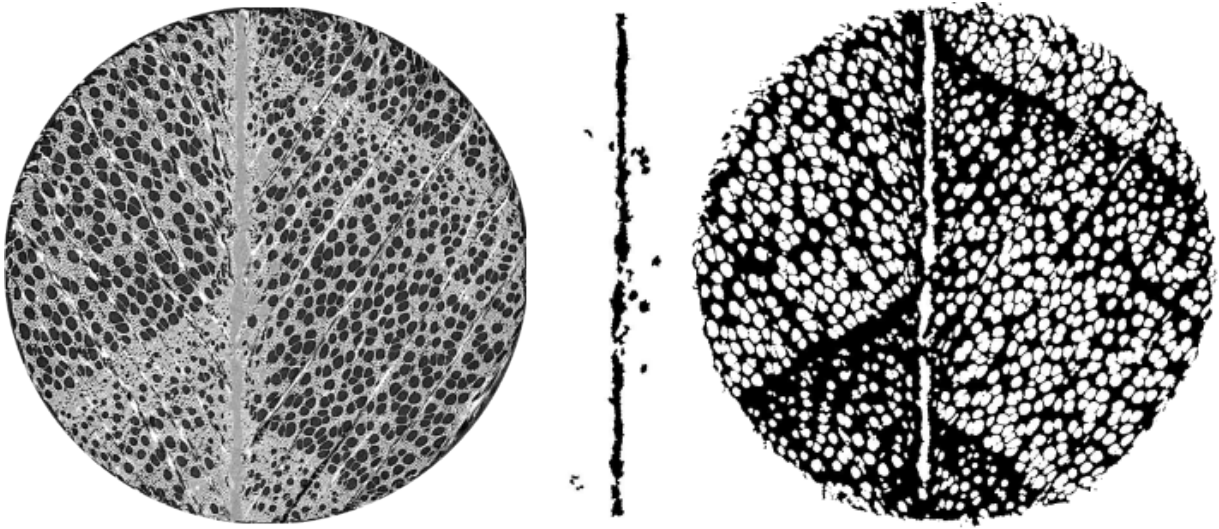

**Figure 5:** CT cross section, segmented adhesive, and pore space.

By interpolating the picked gray scale values over the whole volume, we can extract the bond lines from the samples. Finally we can create iso-surfaces of the identified adhesive for visualization purposes (Figure 6).

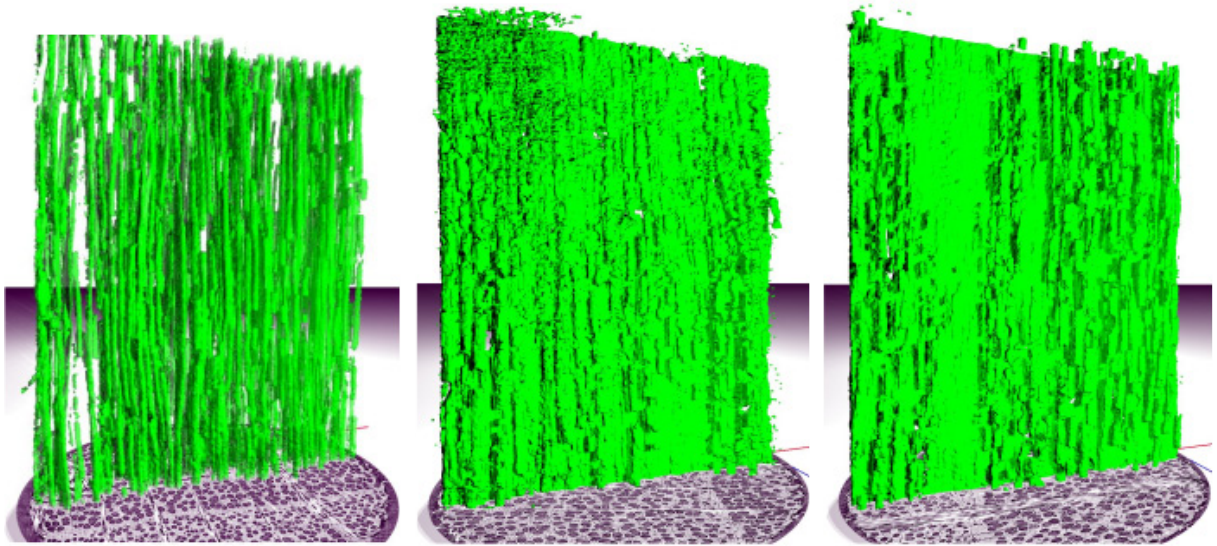

**Figure 6:** Extracted bond lines for PUR-2 (0°), UF-1 (75°), and PVAC-2 (90°) from left to right.

**Wood Ray Distribution**

As proposed in (Hass et al. 2010), the tangential deflection of the vessels by the wood rays is typical for beech wood. To complete the characterization of anatomical features that have an influence on the development of the bond line, the tomographic data is used to evaluate the proportion of wood rays on the total wood material. Therefore, the data sets are aligned in the LT plane and the wood rays are identified by digital image analysis (Figure 7).

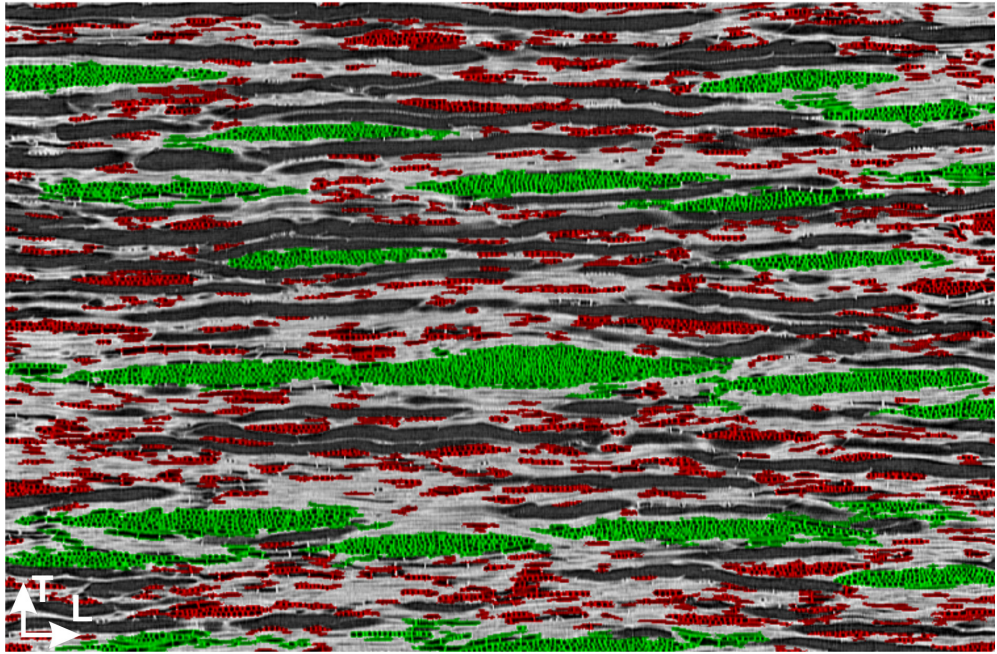

**Figure 7:** Identified wood rays in beech wood in the LT-plane located in early wood. The 20 largest segmented rays are colored in green, all other ones in red. The portion of rays is 24% and the identified grain angle was 4.3°.

# Results

Adhesives penetration into the porous structure occurs following the path of least resistance, either by gross penetration or cell-wall penetration (Kamke and Lee 2007). We observe the filling of accessible, cut cell lumen and mostly the interconnected network of vessels, and the open space between the two contacting rough surfaces. Cell-wall penetration as well as penetration through pits and small cell lumen was beyond the scale of observation. The flow is initiated by capillary suction and the applied pressure. However the local magnitude of the pressure in the volume of the sample remains unknown in our experiments due to surface roughness and the strong heterogeneity and meso-structure of the wood itself. The gross penetration in beech wood is dominated by the vessel network characteristics like vessel diameter, density, connectivity and alignment as well as the temporal and spatial evolution of the viscosity of the adhesives.

When comparing the adhesive systems, even though the initial viscosity of PUR is comparable to the ones of PVAC and UF, its penetration is far deeper (Figure 8).

**Figure 8:** Cross sections of selected samples. The height of the white bar corresponds to 1mm. Note that gray values of the adhesives differ, depending on their density and composition. Abbreviations: SoNA → side of non-application; SoA → side of application.

This is mainly due to the fact, that a pure prepolymer without fillers and additives was used. The viscosity change is only due to polymerization and takes place long after the penetration. The fact that the penetration is rather fast can be seen when comparing the application side with the other beech wood part. As a result, bond lines starve, which is quite typical for the PUR-prepolymers. PVAC exhibits a similar penetration behavior as UF showing a distinct bond line and only minor penetration into the pore structure, mainly filling all directly accessible volume including cut open vessels. The penetration of contacting vessels via pits however is rarely observed.

The influence of the growth ring angle on the penetration pattern is best explained on the extremes of 0° and 90° samples (see Figure 8). For all adhesives, it seems that the penetration is deeper for 90° than for 0°, confirming observations by (Sernek et al. 1999). Basically adhesives mainly penetrate the longitudinally aligned vessels. However as shown by (Bosshard 1973; Hass 2010), vessels in beech wood have a strong waviness in tangential direction since they have to weave around the radially directed wood rays. Therefore, looking at cross sections, it is obvious that the impression of a deeper penetration in tangential direction is created. Another reason for this impression is the anatomical characteristic of beech wood to build a growth ring limited vessel network (Bosshard 1973; Hass 2010), which causes the growth ring border to act as a barrier for the penetration. Therefore, no vessels outside the cut growth ring are filled (Figure 8 left). At a GRA of 90° all growth ring borders are oriented perpendicular to the bond line, while at a GRA of 0° they lie parallel to it, blocking further advancing of the adhesive away from the bondline.

The established way to characterize the penetration behaviour of an adhesive is the determination of the maximum penetration depth in single cross sections. As this might be a feasible method for the penetration in softwoods, it seems not appropriate for hardwoods such as beech wood, where strongly different bond line morphologies can be observed (see comparison Figure 1). Therefore we propose the saturation of the available pore space by the adhesive for characterization, which represents the ratio of the amount of adhesive and the porosity of the sample without the adhesive. In Figure 9 the two ways of determining the adhesive penetration are exemplarily demonstrated on a UF bond line.

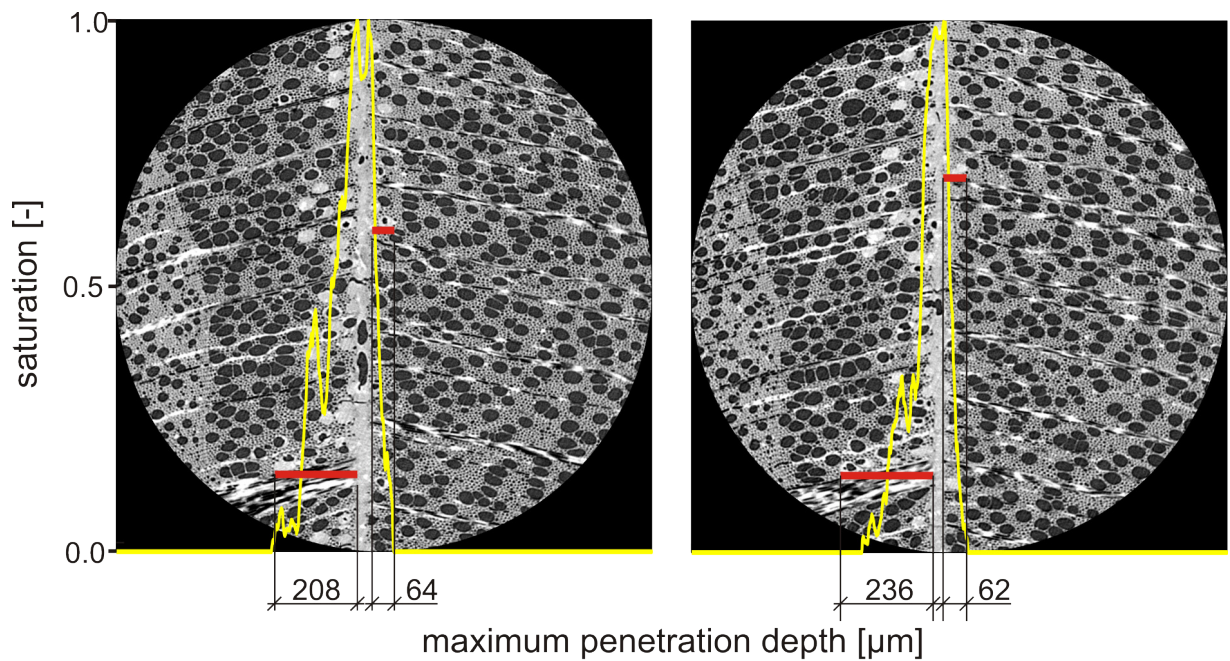

**Figure 9:** Saturation and maximum penetration depth in two cross-sections of a specimen containing a bond line of an UF-adhesive. The distance in longitudinal direction between the cross-sections is 100 slices or 370µm.

Several imponderabilities of the usual investigation methods can be recognized. On the one hand, the maximum penetration depth lacks important information about the bond line morphology, which applies not only on the penetration itself, but as well on the pure bond line between the two adherents. Hence imperfections in the bond line, which will have a huge impact on the strength of the bonding, are not taken into account, but isolated adhesive filled pores, whose contribution to the bonding strength is questionable, define its value. Those bond line imperfections are developed in an adhesive characteristic way: For PUR bond line starvation was typical due to high mobility of the adhesive. For PVAC, that behaves rather ductile, we observe pore formation inside the bulk adhesive due to excessive shrinkage, while for UF typical crack patterns due to restrained shrinkage of the brittle adhesive appear. Typical examples are shown in Figure 10.

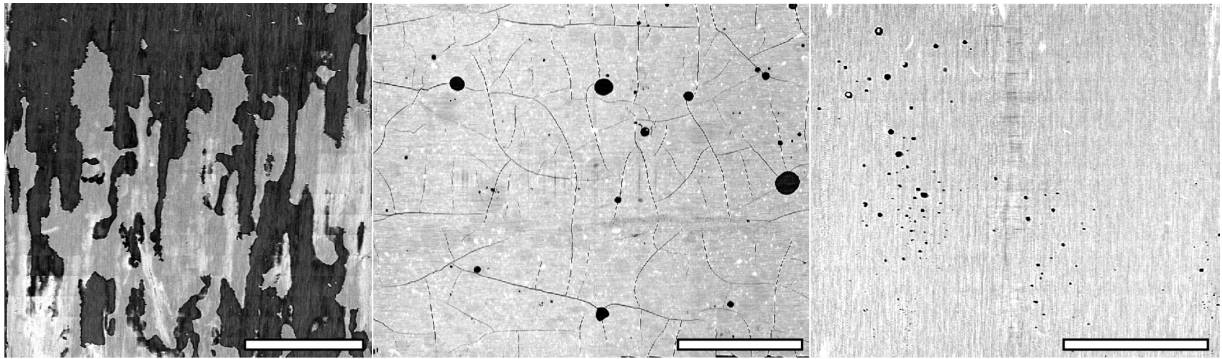

**Figure 10:** Bond line imperfections from left to right: Starved PUR, cracked UF, and PVAC with multiple voids. The width of the white bar corresponds to 1mm.

On the other hand, the example shown in Figure 9 demonstrates how the results of the measurement depend on the investigated cross-section and how the bond line morphology change along the longitudinal axis. Although the two cross-sections are only 370µm distant, the maximum penetration depth on the left hand side is about 15% smaller and contains two big holes in the adhesive layer. Therefore the information available from single cross-sections is rather limited, while several investigations on the same bond line give a more detailed description about the bond line characteristics.

First we average over all samples per adhesive, to demonstrate the differences between them. Due to the adhesive application procedure, the penetration characteristics of both sides of the PVAC and UF bondings can be seen as repetitions with identical conditions. This way the mean saturation distribution for an adhesive was evaluated until the middle of the bond line was reached and mirrored for demonstration reasons. As the application of the PUR adhesives was carried out single-sided, a differentiation between side of application (SoA) and side of non-application (SonA) (see Figure 8 left) had to be made. Therefore the simplification as for UF and PVAC could not be made. In Figure 11, the mean values for both parameters per adhesive are presented.

**Figure 11:** Mean values for saturation and maximum penetration depth (inset). Abbreviations for Adhesives in inset: a → PVAC-1; b → PVAC-2; c → UF-1; d → UF-2; e → PUR-1, side of application (SoA); f → PUR-1 side of non-application (SoNA); g → PUR-2, SoA; h → PUR-2, SoNA.

Figure 11 illustrates that the maximum penetration depth shows no distinct difference between UF and PVAC. The saturation however indicates that the UF adhesives develops a bond line where the adhesive layer shows less voids, while the PVAC-bond line inherit more areas with no adhesive at all like holes and pores. Between the two PVAC adhesives, the saturation shows differences as well. If we compare the saturation per adhesive with the results of the lap-shear tests in (Hass et al. 2009), we discover the same relations for the shear strength as for the saturation: For the UF adhesives, no difference between the two viscosity steps could be detected, while for the PVAC adhesives, a significant influence of the adhesive viscosity could be found. Surprisingly, although in the PVAC-1 bond lines more voids could be detected, higher shear strength was reached with these specimens compared to those bonded with PVAC-2. It is possible, that pores lead to crack arrest due to crack tip blunting, but more research on the micro-mechanical failure behavior of wood bondings is necessary to clarify these questions. The saturation also delivers the explanation for the low wood fracture percentage of the PUR bondings in (Hass et al. 2009): As the low saturation with these adhesives reveals, the penetration is so excessive, that the bond line partially starves, providing a weak area of preferred failure.

In a next step we look at the overall volume of segmented adhesive. For UF and PVAC, the volumes are as expected. For PUR, although the chemical structure of the two PUR prepolymers was comparable and the applied quantity of adhesive was equal, we observe a difference in the amount of hardened adhesive in the bondings. We detect more of the prepolymer-2 with the higher cross-linking density and the higher viscosity. This can be the result of at least two scenarios: Due to the higher cross-linking density, the hardening reaction of the prepolymers can differ, leading to a higher amount of $CO_2$ and a higher volume of the hardened adhesive. Another reason could be the difference in the chain length distribution of the two adhesives. As the prepolymer-1 with the lower cross-linking density inherits a higher fraction of short chains, a penetration into the call walls becomes more likely. This kind of penetration is not detectable within this investigation and thereby excluded.

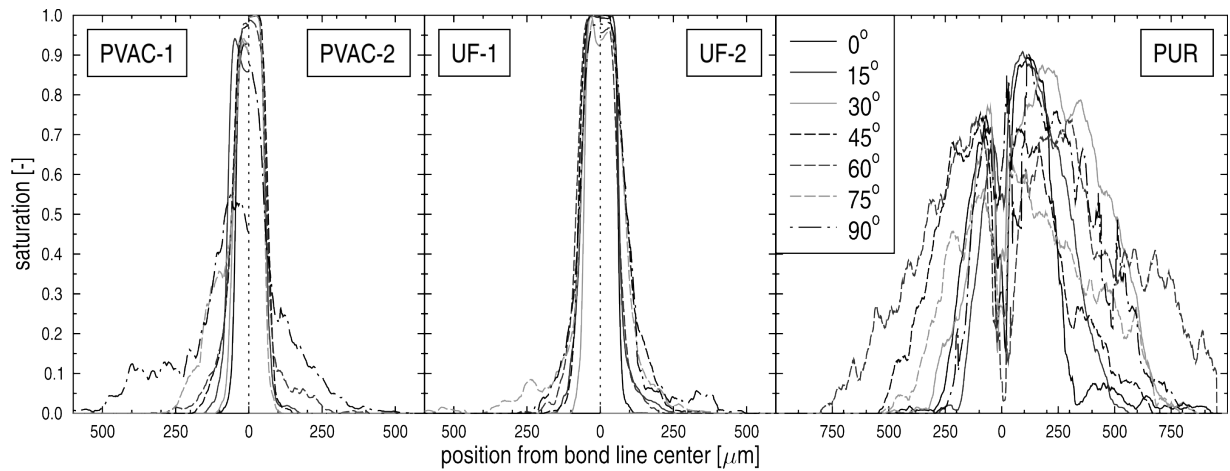

**Figure 12:** Saturation profiles for all adhesive systems and growth ring angles. Note that each curve is an average of 9 different regions inside one sample. PUR curves are rotated in a way that the application side is always on the right.

The dependence on the year ring angle can be identified by confronting the various saturation profiles. As noted before, the sample size does not allow for a profound statistical evaluation. However we observe the trend that PVAC and UF exhibit an increasingly wider profile for larger growth ring angles. For PUR however the widest saturation profiles can be found at growth ring angles between 45° and 60° (see Figure 12).

## Conclusions

We demonstrated that the characterization of adhesive bond lines in wood by the analysis of single cross-sections in the RT-plane and determination of the maximum penetration depth lacks important information about the bond line morphology. For bond lines in softwood, this method might be sufficient, since the bond line develops a more or less interconnected and evenly distributed penetration area. In hardwoods however, the bond line strongly depends on the anatomical characteristics of the wood species, as the main penetration occurs in the vessel network. Therefore we propose the saturation of the available pore space as a parameter, which is physically more meaningful. With this method imperfections of the bond line are considered as well as the filling of single, isolated vessels is neglected, because they have only minor influence on the bonding strength. We exemplarily showed the suitability of the saturation, as we were able to relate findings from a previous investigation on bonding shear strength to it, where the maximum penetration depth allowed no relation to the mechanical performance.

The usage of three-dimensional data allowed us to investigate the development of the bond line morphology along its course. Thereby it became clear, that the bond line characteristics strongly depend on the measuring point. Therefore the investigation of only one cross-section per sample gives only limited information about the bond line. The volumetric data also provided a look in the cured bond line presenting several imperfections which are characteristic for the different adhesive types.

With the identification of the vessel network in (Hass 2010), the wood ray distribution and the determination of the adhesive hardening and penetration behavior, all input parameters are available to setup a first model of hardwood penetration by hardening adhesives in part II of this work.

## Acknowledgements

The financial support of this work under SNF grant 116052 is acknowledged. Furthermore we are thankful for the access to the SLS beam line TOMCAT and its crew, namely S.A. McDonald, F. Marone, M. Stampanoni and the valuable advices on data evaluation by A. Kästner.